\documentclass{emulateapj}
\usepackage{apjfonts}
\usepackage{epsf}
\usepackage{psfig}
\usepackage{graphicx}
\begin{document}

\slugcomment{Submitted to ApJ}
\shortauthors{J. M. M\"iller \& K. G\"ultekin}
\shorttitle{Swift J164449.3$+$573451}

\title{X-ray and Radio Constraints on the Mass of the Black Hole in
  Swift J164449.3$+$573451}

\author{J.~M.~M\"iller\altaffilmark{1},
  K.~G\"ultekin\altaffilmark{1},
}

\altaffiltext{1}{Department of Astronomy, University of Michigan, 500
Church Street, Ann Arbor, MI 48109, jonmm@umich.edu}

\begin{abstract}
Swift J164449.3$+$573451 is an exciting transient event, likely
powered by the tidal disruption of a star by a massive black hole.
The distance to the source, its transient nature, and high internal
column density serve to complicate several means of estimating the
mass of the black hole.  Utilizing newly-refined relationships
between black hole mass, radio luminosity, and X-ray luminosity, and
de-beaming the source flux, a weak constraint on the black hole mass
is obtained: log$(M_{BH}/M_{\odot}) = 5.5 \pm 1.1$ (1$\sigma$
confidence).  The confidence interval is determined from the current
intrinsic scatter in the relation, which includes effects from X-ray
variability and accretion modes.  This mass range is broad,
but it includes low values that are consistent with some variability
arguments, and it safely excludes high mass values where it becomes
impossible for black holes to disrupt stars.  Future refinements in
relationships between black hole mass, radio luminosity, and X-ray
luminosity will be able to reduce the uncertainty in related mass
estimates by a factor of two, making this technique comparable to
estimates based on the $M-\sigma$ relationship.  Possible difficulties
in placing such events on the fundamental plane, a potential future
test of their suitability, and uncertainties in mass stemming from
variable X-ray emission are discussed.  As near and longer-term survey
efforts such as Pan-STARRS, LSST, LOFAR, the Square Kilometer Array,
and eROSITA begin to detect many tidal disruption events, black hole
mass estimates from combined X-ray and radio observations may prove to
be very pragmatic.
\end{abstract}

\section{Introduction}
Swift J164449.3$+$573451 was orginally detected on 28 March 2011 as a
gamma-ray burst, GRB 110328A, via the {\it Swift} X-ray observatory.
The properties of the source, including its variability and longevity,
quickly made clear that the source must be a new kind of transient
event.  Optical observations soon revealed a red-shift of $z=0.354$
for the host galaxy (Levan et al.\ 2011), and a position close to the
galactic center.  Based on the source properties, including variations
in the X-ray spectrum and flux, it was suggested that the transient
source may be powered by the tidal disruption of a star by a
supermassive black hole (Bloom et al.\ 2011), viewed close to the axis
of a jet.  Subsequent and detailed studies support this interpretation
(e.g. Bloom et al.\ 2011b, Burrows et al.\ 2011, Levan et al.\ 2011b,
Krolik \& Piran 2011).

The mass of the black hole that likely disrupted the star is an
intersting and important question.  Rees (1988) predicts
the disruption of normal stars is possible for black holes for masses
below $10^{8}~ M_{\odot}$ and describes observational hallmarks of the
aftermath.  Changes in the gravitational force across the diameter of
a star are too gentle for black holes above this mass range, and stars
are then swallowed whole.

In the case of Swift J164449.3$+$573451, its distance makes it
impossible to trace stellar orbits close to the compact object and to
establish a mass by resolving the dynamical sphere of influence (see,
e.g., G\"ultekin et al.\ 2009).  A mass via the $M$--$\sigma$
relationship is possible but may not be easy given the source
distance, flux, and unknown host morphology.  Other methods for
estimating black hole masses are tied to direct primary masses and the
$M$--$\sigma$ relationship, such as reverberation mapping
(e.g. Peterson et al.\ 2004).  However, this may also be complicated
by strong absorption along the line of sight, and by the transient
nature of the source.

A new method for estimating black hole masses relies on the
``fundamental plane'' of black hole accretion (e.g. Merloni, Heinz, \&
Di Matteo 2003; Falcke, Kording, \& Markoff 2004).  The plane reveals
a relationship between radio luminosity, X-ray luminosity, and
black hole mass.  Radio luminosity serves as a proxy for jet power,
and X-ray luminosity for accretion power.  The ability of the plane to
predict a black hole mass based on radio and X-ray luminosity
estimates was sharpened by G\"ultekin et al.\ (2009b), wherein only
black holes with direct primary masses are used.  Inverting the
plane to give masses shows great promise: current mass
estimates are only a factor of two less certain than estimates from
the $M$--$\sigma$ relationship itself, and the addition of more
sources and simultaneous radio and X-ray observations may make it a
comparably reliable predictor.  The plane was recently used to
estimate the mass of the central black hole in the dwarf starburst
galaxy Henize~2-10 (Reines et al.\ 2011).

It has been unclear how to treat beamed sources, with
respect to the fundamental plane.  The power of the plane is that it
gives information on the central engine -- how accretion inflow and
relativistic jet outflows are coupled.  Jetted sources, in contrast,
may not allow a clean view of the central engine.  Plotkin et
al.\ (2011) have recently shown that beamed sources -- including BL
Lac objects -- fall on the fundamental plane once their fluxes are
de-beamed.  This methodology may not be uniformly applicable to other
sources, but it demonstrates that beamed sources follow the same
relation as others once a careful analysis of the relativistic effects
is made.  Coupled with the ability of the refined plane to give mass
estimates, such treatments enable an estimate of the black hole mass
in sources such as Swift J164449.3$+$573451.  

In the sections that follow, we detail a mass estimate for Swift
J164449.3$+$573451 based on simultaneous radio and X-ray observations
and fundamental plane relations.  This estimate requires several
assumptions; possible drawbacks and a possible test of whether or not
such sources can be placed on the fundamental plane are included in
Section 4.

\section{Data Selection and Reduction}
To place Swift J16449.3$+$573451 on the fundamental plane of black
hole accretion, radio and X-ray luminosity points are required.
The importance of obtaining contemporaneous radio and X-ray points in
drawing physical inferences from the plane, and in reducing its
internal scatter, is becoming clear (e.g. King et al.\ 2011, Jones et
al.\ 2011).  We therefore selected publicly available fluxes
separated by the smallest possible margin:

Bower, Bloom, \& Cenko (2011) observed Swift J16449.3$+$573451 using
the VLBA on 1 April 2011, starting at 05:30 UT.  They measure a flux
density of $1.7 \pm 0.1$~mJy at 8.4~GHz, with an upper limit of just
17\% on variations in the flux density within the four hour
observation.  In a more recent analysis, Zauderer et al.\ (2011)
report a flux density of 0.82(2) mJy at 4.9~GHz on April 1.27 (6:48
UT).  The fundamental plane is constructed using flux density
measurements in different bands, shifted to 5~GHz.  Also on 1 April
2011, the {\it Swift}/XRT observed the source in PCW3 mode for a total
of 21 ks, starting at 00:52 UT.  Thus, both the 8.4 GHz and 4.9 GHz
radio measurements had brief periods of overlap with this early X-ray
observation.

The XRT data were reduced using the latest HEASOFT suite (version
6.10) and a fully up-to-date CALDB.  Source events were extracted from
the cleaned PCW3 event list in an annular region centered on the
source, with an inner radius of 10 pixels and an outer radius of 40
pixels.  This region was chosen to prevent photon pile-up distortions
to the spectrum and flux.  Background events were extracted from a
circular region with a radius of 20 pixels, located well away from the
source.  The tool "xselect" was used to create source and background
spectra.  The standard redistribution matrix file from the calibration
database was used, and an ancillary response function was generated
using the tool "xrtmkarf".

The spectra were grouped to require at least 10 counts per spectral
bin using the tool "grppha", and the spectrum was fit using XSPEC
version 12 (Arnaud \& Dorman 2000).  The spectrum below 0.5 keV and
above 10.0 keV (in the observed frame) was ignored owing to
calibration uncertainties in those bands.  Neutral absorption along
the line of sight was fit using the "ztbabs" model (Wilms, Allen, \&
McCray 2000), which places photoelectic absoprtion edges at the energy
appropriate given the red shift of the source.  This simple model
gives an adequate fit, $\chi^{2}/\nu = 620.8/529 = 1.17$.  A high
column density is measured, ${\rm N}_{\rm H} = 1.32(4)\times 10^{22}~
{\rm cm}^{-2}$.  The power-law photon index is measured to be $\Gamma
= 1.73(3)$; this slope is canonical for Seyfert-1 AGN (see, e.g.,
Nandra et al.\ 1997).  A power-law flux normalization of $9.1(3)
\times 10^{-3}~ {\rm ph}~ {\rm cm}^{-2}~ {\rm s}^{-1}$ is obtained.
This gives a 2--10~keV (observed frame) flux of $3.4(1) \times
10^{-11}~ {\rm erg}~ {\rm cm}^{-2}~ {\rm s}^{-1}$, corresponding to a
2--10~keV flux of $2.9(1) \times 10^{-11}~ {\rm erg}~ {\rm cm}^{-2}~
{\rm s}^{-1}$ in the emitted frame.  All of the above errors
correspond to $1\sigma$ uncertainties.

\section{Mass Estimation}
Although the flux density measurement at 8.4~GHz (Bower, Bloom, \&
Cenko 2011) has move overlap with the X-ray observation employed in
this work, the flux density of 0.82(2) mJy at 4.9~GHz reported by
Zauderer et al.\ (2011) is more relevant.  The FP of Merloni, Heinz \&
Di Matteo (2003), for instance, was constructed by gathering radio
core measurements at different frequencies and shifting them to 5~GHz.
The FP contains a broad range of sources, from LINERs with core
emission that can be optically thick to Seyferts with
emission that is optically thin (for a review of AGN core properties,
see, e.g., Ho et al.\ 2008). 

The sources in FP relations are typically closer than Swift
J164449.3$+$573451, and in order to use the FP, then, we need $L_R =
\nu L_\nu$ at 5 GHz in the observed frame.  Zauderer et al.\ (2011)
report on a number of radio observations on 30 March 2011 and find
$F_\nu \propto \nu^{1.3\pm0.1}$.  Based on their tables, we confirmed that
  the same index holds on 1 April 2011, and shifted the flux density
  to 5~GHz in the observed frame for this assumed spectrum.   At a
  redshift of $z = 0.3534$ (Levan et al.\ 2011b) the luminosity
  distance is $D_L = 1.8 \mathrm{Gpc}$ for an assumed cosmology of $h
  = 0.71$, $\Omega_m = 0.27$, and $\Omega_\Lambda = 0.73$.  Thus
  assuming isotropic emission, $L_R = 8.26 \times
  10^{39}\ \mathrm{erg\ s^{-1}}$.  For the same distance, the
  restframe 2--10 keV flux corresonds to an isotropic luminosity of
  $L_X = 1.12 \times 10^{46} \mathrm{erg\ s^{-1}}$.

In the case of Swift J164449.3$+$573451, observations of radio
scintillation by Zauderer et al.\ (2011) are helpful: a
Lorentz factor of $\Gamma \simeq 5$ is implied by their measurements.
We have assumed this value of $\Gamma$ in de-boosting the emission
observed from Swift J164449.3$+$573451.  Special care is
needed: the specific de-boosting calculations employed by Plotkin et
al.\ (2011) assume optically-thin emission (also see Lind \& Blandford
1985) and may not appropriate for Swift J164449.3$+$573451, or at
least not at early times.  An expression for beaming due to
relativistic motions for a generic spectral index is given by: $F =
F_0 (1-\beta\mu)^{-(2+\alpha)}$ (where $F$ is the flux observed, $F_0$
is the flux emitted in the source frame, $\beta = v/c$ and derives
from $\Gamma$, $\mu = cos(\theta)$, and $\alpha$ is the spectral index
of the emitter; see, e.g., Peacock 1999).  Assuming that $\mu = 1$, we
derive a correction factor of 15.5.  The intrinsic
luminosities are then $L_R = 5.3 \times 10^{38}\ \mathrm{erg\ s^{-1}}$ and
$L_X = 7.2 \times 10^{44}\ \mathrm{erg\ s^{-1}}$.

Although many FP fits exist for different source collections and aims
(e.g. Merloni, Heinz, \& Di Matteo 2003; Falcke, Kording, \& Markoff
2004; Yuan \& Cui 2005; G\"ultekin et al.\ 2009b, Plotkin et
al.\ 2011),we use the FP from G\"ultekin et al.\ (2009b), which was fit using
only sources with dynamically determined black hole masses.
G\"ultekin et al.\ (2009b) gave several fits using different samples
and statistical methods.  We use the fit defined by their equations 1
and 10, which has the same functional form as fits by Merloni et al.\
(2003) and includes stellar-mass black holes.  We chose this version
because it should interpolate between the $10^{7}$ and $10^{1}\
M_{\scriptscriptstyle \odot}$ black holes.  The mass predictor fit is
only valid in its fitting domain of $M > 10^{6}\
M_{\scriptscriptstyle \odot}$, and $L_X < 10^{43}\ \mathrm{erg\
s^{-1}}$.  The fit we are using can be written as
\begin{equation}
\log L_R = (6.31 \pm 0.21) + (0.82 \pm 0.08) \log M + (0.62 \pm 0.10) \log L_X,
\label{eqradiofp}
\end{equation}
where $L_R$ and $L_X$ are in units of $\mathrm{erg\
s^{-1}}$ and $M$ is in solar mass units.  The rms
intrinsic scatter of this relation is 0.88 dex in the $\log L_R$
direction.  This may be inverted to write
\begin{equation}
\log M = 1.22 (\log L_R \pm 0.88) - 0.76 \log L_X - 7.70,
\label{eqfp}
\end{equation}
dropping coefficient uncertainties since the scatter dominates.
Using this relation with the luminosities above, we find $\log M = 5.5
\pm 1.1$, corresponding to $M = 3.2 \times 10^{5}\ M_{\scriptscriptstyle
\odot}$ with a 68\% confidence range of $0.25$--$40 \times 10^{5}\
M_{\scriptscriptstyle \odot}$.  In Figure 2, we show  Swift
J164449.3$+$573451 on the FP with SMBHs and stellar-mass BHs.

\begin{figure}
\includegraphics[angle=-90,width=\columnwidth]{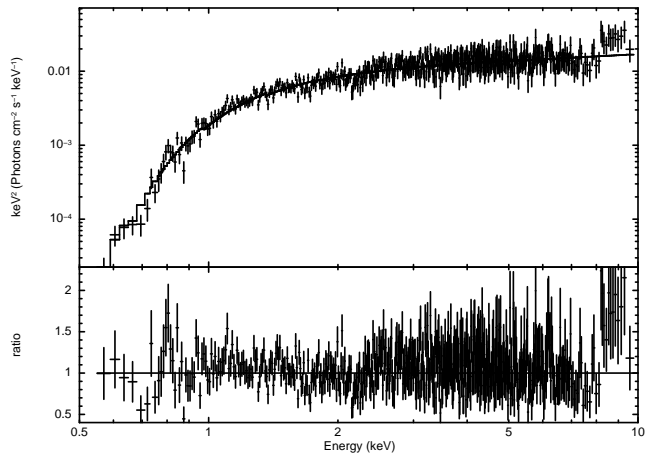}
\figcaption[h]{\footnotesize The figure above shows the unfolded XRT
  spectrum of Swift J164449.3$+$573451, obtained on 1 April 2011.  The
  top panel shows the unfolded spectrum, and the bottom panel shows
  the ratio of the data to a simple absorbed power-law model  with $\Gamma
  = 1.73(3)$.  The turn-over at low energy is the result of a high
  column density, ${\rm N}_{\rm H} = 1.32(4) \times 10^{22}~ {\rm
    cm}^{-2}$.  This basic spectral model allows for a good
  characterization of the total X-ray flux in the XRT band.}
\end{figure}
\medskip

We may compare this mass estimate to a prediction based on the
relation between black hole mass and the host galaxy's bulge
luminosity, the $M$--$L$ relationship.  In the $V$-band, it is given
by $\log M = (8.95 \pm 0.11) + (1.11 \pm 0.18) \log(L_V / 10^{11})$,
where $L_V$ is given in units of $V$-band solar luminosities and the
rms intrinsic scatter is $0.38 \pm 0.09$ dex (G\"ultekin et al.\
2009).  The host galaxy of Swift J164449.3$+$573451 has $L_B =
10^{9.20}$ (Leloudas et al.\ 2011; Burrows et al.\ 2011), which is very close to restframe
$V$-band.  Thus the mass predicted by the $M$--$L$ relation is $\log M
= 6.95 \pm 0.38$, corresponding to $M = 9 \times 10^{6}\
M_{\scriptscriptstyle \odot}$ with a 68\% confidence interval of
$3.7$--$21 \times 10^{6}\ M_{\scriptscriptstyle \odot}$.  The $H$-band
luminosity of $L_H = 10^{9.58}$ is very close to restframe $K$-band
and predicts (Marconi \& Hunt 2003) a black hole mass of $\log M =
6.72 \pm 0.31$, corresponding to $M = 5.2 \times 10^{6}\
M_{\scriptscriptstyle \odot}$ with a 68\% confidence interval of
$2.6$--$11 \times 10^{6}\ M_{\scriptscriptstyle \odot}$.  As noted by
Burrows et al.\ (2011), any black hole mass inferred from the $M$--$L$
relation for Swift J164449.3$+$573451 is an upper limit because this
is the total luminosity of the galaxy, not the bulge.  The spectrum of
the galaxy indicates that it is star-forming and therefore a spiral
galaxy.  Since bulge mass is a much better predictor of black hole
mass than total luminosity (Kormendy, Bender, \& Cornell 2011), we can only put an
upper limit on the bulge mass and thus the black hole mass.  If the
galaxy is bulgeless, it is possible that the black hole mass is much
smaller than would be inferred by assuming that the $M$--$L$ relation.

\section{Discussion}
We have presented a weak but purely observational constraint on the
mass of the black hole in the candidate tidal disruption event Swift
J164449.3$+$573451.  De-beaming simultaneous radio and X-ray fluxes,
and exploiting the ability of refined versions of the fundamental
plane of accretion to estimate black hole masses, we find
log$(M_{BH}/M_{\odot}) = 5.5 \pm 1.1$.  This constraint is compatible
with the predictions of Rees (1988).  The allowed range includes more
extreme but exciting interpretations of the event, including the
disruption of a white dwarf by a black hole with a mass below $10^{5}~
M_{\odot}$ 

\begin{figure}[thb]
\includegraphics[width=\columnwidth]{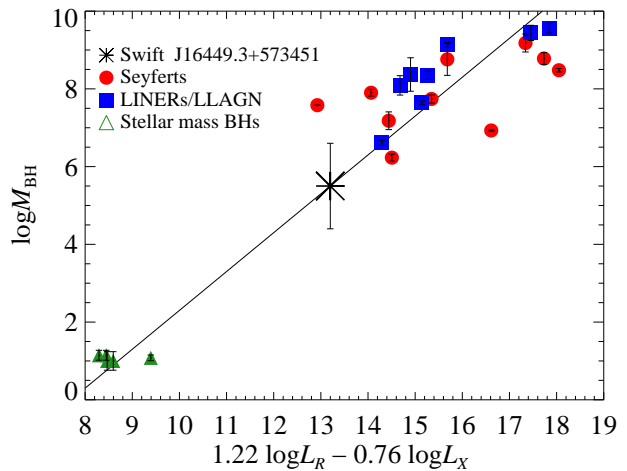}
\figcaption{The inverted fundamental plane of black hole accretion.  The
sources with measured masses are from G\"ultekin et al.\ (2009b), The
position of Swift J164449.3$+$573451 is shown with a cross.  With the
assumption that the FP holds for this new and interesting source, the
observed X-ray and radio luminosities imply that the black hole has
mass $\log M = 5.5 \pm 1.1$.}
\end{figure}
\medskip

\noindent (Krolik \& Piran 2011), and it excludes black hole masses
that are so large that stars would simply be accreted whole.
Tentative evidence of an X-ray quasi-periodic oscillation (QPO) 
in Swift J164449.3$+$573451 (Miller \& Strohmayer 2011)
implies a mass of log$(M_{BH}/M_{\odot}) = 5.7$, assuming the
oscillation reflects the Keplerian orbit at the innermost stable
ciruclar orbit around a Schwarzschild black hole.  It is notable that
the central value of the plane--derived mass and the mass implied by the
potential QPO are very close.

In this analysis, the source flux was de-beamed using a simple and
well-established relationship.  This effort benefitted from a
recent approximation of the Lorentz factor based on radio
scintillation (Zauderer et al.\ 2011).  Nevertheless, de-beaming any
observed flux is difficult, and prone to uncertainties.  The method 
employed is general in that it can be adapted to any instrinsic source
spectral shape; however, difficulties enter when sources are optically
thick, with the potential effect of broadening the beaming cone beyond
$\theta \simeq 1/\Gamma$ (e.g. Lind \& Blandford 1985).  If this is
important in the case of Swift J164449.4$+$573451, then the (modest)
correction factor may be too high and the mass may be biased to low
values.  Omitting any beaming correction, our results imply a mass of
log$(M_{BH}/M_{\odot}) = 6.0 \pm 1.1$; the upper limit is again within
the range where massive black holes can disrupt stars.

Additional difficulties enter in that the early X-ray flux of Swift
J164449.3$+$573451 was strongly variable, whereas the radio flux
density was comparatively stable (e.g. Levan et al.\ 2011b, Zauderer
et al.\ 2011).  In this and other sources, variability can lead to
uncertainties in the mass.  Indeed, the non-simultaneity of many
points on the fundamental plane may be an important source of scatter.
This analysis attempts to limit related uncertainties by selecting
data with at least small periods of strict simultaneity.  Moreover, it
is worth noting that early spectroscopy by Bloom et al.\ (2011) report
possible variations in the column density and the possible presence of
a blackbody in the spectrum.  These may serve to indicate optical
depth variations.  The column density and simple power-law spectrum
measured in the {\it Swift}/XRT spectrum used in this analysis are
similar to those in later deep observations with {\it XMM-Newton}
(Miller \& Strohmayer 2011).
 
A major assumption of our work is that the tidal disruption and
accretion of a star by a massive black hole emits radiation as an
accretion flow that bears strong similarities to a standard disk,
corona, and jet accretion flow geometry that is widely thought to hold
in AGN.  This is a strong assumption, but it can be tested with future
simultaneous X-ray and radio observations.  The tentative detection of
an X-ray QPO in Swift J164449.3$+$573451 (Miller \& Strohmayer 2011)
may provide some support for this assumption.  Because of the rapid
variability of this source compared to typical variability timescales
for SMBHs, simultaneous X-ray and radio observations in
multiple-epochs and spanning a decade or more of variation in each parameter
would reveal whether or not $L_R \sim L_X^{0.7}$ (e.g. Gallo, Fender,
\& Pooley 2003).  If so, it would strongly suggest that accretion
energy from tidal disruption events is radiated in the same way as
typical accretion disks and thus validate our approach.

Black hole mass estimates for Swift J164449.3$+$573451 based on
flaring timescales require assumptions about how the timescale
associates with orbital timescales, or other timescales in the system,
although many of the arguments are compelling (e.g., Krolik \& Piran
2011).  Unless the luminosity of host galaxy bulges can be reliably
constrained, black hole mass estimates based on the $M$--$L$
relationship will only give upper limits.  The FP mass estimate range
of two orders of magnitude is the best observational constraint one
can presently give on the black hole in Swift J164449.3$+$573451, but
this could be improved upon with future data and scaling relations.

Currently, the FP is based on only 18 black holes with
dynamically-measured masses.  The rest come from reverberation
mapping or secondary scaling relations, such as the $M$--$\sigma$
relation.  The uncertainty in predicting mass is dominated by the
intrinsic scatter of the current plane, for which several causes have
been identified: (1) X-ray and radio observations from vastly
different epochs (G\"ultekin et al.\ 2009b; King et al.\ 2011); (2)
heterogeneous observations (observatories, bands, modes, and analysis
protocols; G\"ultekin et al.\ 2009b); (3) potentially distinct
accretion modes (G\"ultekin et al.\ 2009b; King et al.\ 2011; Plotkin
et al.\ 2011); and (4) uncorrected beaming of radiation (Plotkin et
al.\ 2011).  

We are now engaged in a joint \emph{Chandra} and EVLA program to
survey all SMBHs with dynamically--measured masses.  Once complete, the
increased numbers ($\sim60$ sources) will allow us to reduce the
effects of the first three sources mentioned above.  (The final source
of scatter, beaming, has been taken into account for Swift
J164449.3$+$573451.)  We estimate that the full sample will reduce the
uncertainty in mass to about 0.5 dex, comparable to the $M$--$\sigma$
and $M$--$L$ relations (G\"ultekin et al.\ 2009).  If a high-quality
absorption spectrum can be obtained for the host galaxy of Swift
J164449.3$+$573451, then an $M$--$\sigma$-based mass is possible.  As
we noted earlier, without knowing the relative contribution of bulge
and disk components, the mass estimate may be too high (Jardell et al.\ 2011).

More broadly, it is worth noting that masses derived using the
fundamental plane may be pragmatic for tidal disruption events.  The
luminosity of these events means that they can be observed from a
broad range of red shifts.  At large distances, it will not be
possible to obtain direct primary masses, and the combination of
distances, timescales, and host galaxy morphology could make it
difficult to obtain standard secondary mass estimates.  X-ray and
radio fluxes are particularly immune to complications such as local
obscuration, and often more immune to confusion than optical and NIR
observations.  New survey efforts, such as Pan-STARRS and LSST, will
likely detect tidal disruption events in far greater numbers than past
efforts.  So too will future radio facilities, such as the Square
Kilometer Array and LOFAR, and even planned X-ray survey missions such
as eROSITA.  Given the observational realities and difficulties of
tidal disruption events, combining radio and X-ray fluxes to derive
black hole masses may prove to be a valuable tool, especially if the
derived masses soon equal $M$--$\sigma$ masses in quality.


\hspace{0.1in}
KG thanks the Aspen Physics Center for their hospitality.  JMM and KG
thank Cole Miller and Doug Richstone for helpful conversations.  We
thank the anonymous referee for helpful comments that improved this
paper.

\end{document}